# Quantification of Trabeculae Inside the Heart from MRI Using Fractal Analysis

Md. Kamrul Hasan[1], Fakrul Islam Tushar[2]
[1,2]Medical Imaging and Applications (M-1)
[1,2]University Center Condorcet - University of Burgundy
[1,2]Le Creusot, France
Email: [1]kamruleeekuet@gmail.com, [2]f.i.tushar.eee@gmail.com

*Abstract*— **Left ventricular non-compaction (LVNC) is a rare cardiomyopathy (CMP) that should be considered as a possible diagnosis because of its potential complications which are heart failure, ventricular arrhythmias, and embolic events. For analysis cardiac functionality, extracting information from the Left ventricular (LV) is already a broad field of Medical Imaging. Different algorithms and strategies ranging that is semi-automated or automated has already been developed to get useful information from such a critical structure of heart. Trabeculae in the heart undergoes difference changes like solid from spongy. Due to failure of this process left ventricle non-compaction occurred. In this project, we will demonstrate the fractal dimension (FD) and manual segmentation of the Magnetic Resonance Imaging (MRI) of the heart that quantify amount of trabeculae inside the heart. The greater the value of fractal dimension inside the heart indicates the greater complex pattern of the trabeculae in the heart.**

*Keywords*— **Left ventricular non-compaction (LVNC); cardiomyopathy (CMP); Left ventricular (LV); Medical Imaging; Trabeculae; fractal dimension (FD); Magnetic Resonance Imaging (MRI).**

## I. INTRODUCTION

Cardiovascular system consists: blood, blood vessel, the heart and lymphatic system. The heart which weighs less than a pound is responsible for controlling the system of blood vessels having a length of over sixty thousand miles and beats 100,000 times a day and more than 2.5 billion times in the average lifetime [1]. According to the World Health Organization (WHO) in 2015, approximately 17.7 million people died from various Cardiovascular deceases (CVDs) which will around 31% of all world deaths [2]. This fact represents how CVDs are affecting human all over the world.

Heart itself is a magical organ. The basic elements of the heart is shown in Fig. 1. Hearts wall is made up of connective tissue, endothelium and cardiac muscle made. The hearts wall has three different layers called epicardium (outer layer), myocardium (middle layer) and endocardium (inner layer). Epicardium act as a protective layer of the heart and directly connected with myocardium which is thickest layer and responsible for cardiac contractions as shown in Fig. 2. The power needed to pump the blood from health to the whole body is generated by left ventricle and it having the thickest myocardium layer. The inner thinner heart layer endocardium which lines the inner hear layer [1]. American Heart Association explained left ventricular (LV) non-compaction as primary cardiomyopathy and it's identified by the thin compacted epicardial layer and thick compacted endocardial myocardial layer in left ventricle [3].

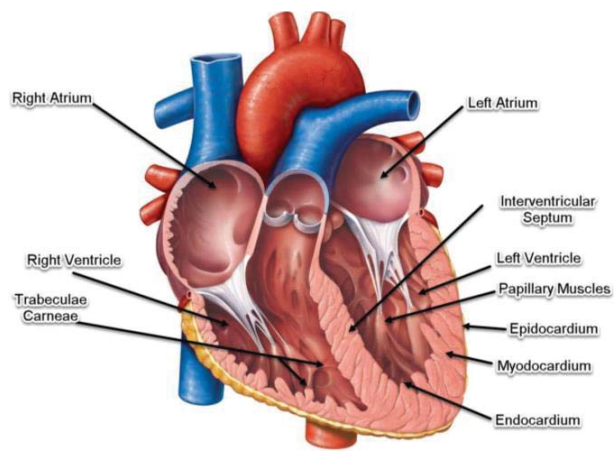

Fig. 1. Typical heart anatomy with its components

But it's not easy to quantify this symptoms of extended trabeculation as normal trabeculae is a feature of normal trabeculation and it also varies from depending on psychical structures. So the most of the existing methods for diagnosis are not fully accurate due not to considering this facts.

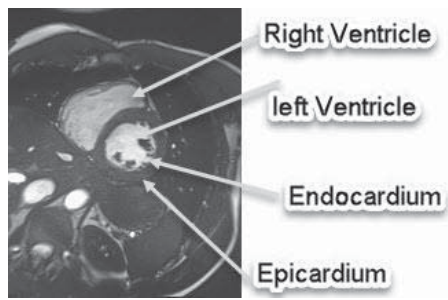

Fig. 2. Different layers of typical heart

According to the given literature fractal analysis could be a possible solution. Fractal analysis is a contemporary method of applying nontraditional mathematics to patterns that defy understanding with traditional Euclidean concepts [4]. Fractal dimension is given by the analysis indicates how completely it fills space [5].

The fractal dimension known as FD indicates the containing information about the target geometrical structure that is a relative measure of complexity. FD is a fundamental analytical parameter which is always a fractional value which quantify how irregular a target pattern is and how much of the space it occupies. The greater the value of fractal dimension the greater the degree of complexity and a more irregular shape of the structure [6]. The box counting method is analogous to the perimeter measuring method where the whole image is to be covered by a grid, and then count how many boxes of the grid are covering part of the image. After that, similar approaches using a finer grid with smaller boxes. In this method of the calculating the FD, the value of FD will be the slope of the straight line of $\log(N)$ on the Y axis and $\log(r)$ on the X axis [7]. Mathematically, FD will be for a given image-

$$FD = \frac{\log(N)}{\log(r)} \quad (1)$$

Where, N is the number of boxes that cover the pattern, and r is the magnification. The algorithm to calculate the FD [8] is described below-

**Step-1:** Depending on the size of each grid $r = 1/2^j$, where $j = 1,2,\ldots,|R|$

**Step-2:** For all the points in the dataset

**Step-3:** Selecting the cell that matched requirement

**Step-4:** Occupancy counter $C_i$ is incremented

**Step-5:** Summation of the square occupancies calculated $S(r) = \sum_i (C_l)^2$

**Step-6:** $\log(r)$ and $\log(S(r))$ value given an plot

**Step-7:** Finding the slope of the linear part from the plot and return FD of dataset A.

The image segmentation problem is fundamental in the field of computer vision. It is a core component toward e.g. automated vision systems and medical applications. Its aim is to find a partition of an image into a finite number of semantically important regions. The active contour model is one of the most successful variation models in medical image segmentation. It consists of evolving a contour in images toward the boundaries of objects. Its success is based on strong mathematical properties and efficient numerical schemes based on the level set method [9]. For the active contour segmentations of the heart MR images, there are two approaches like ballons and snakes [10]. In this thesis, we used snake approach because balloon model introduces an inflation term into the forces. Active contour models like snake model is based on the edge detection and segmentation of magnetic resonance imaging (MRI), computed tomography (CT), and ultrasound medical imagery in which a novel snake paradigm in which the feature of interest may be considered to lie at the bottom of a potential well. Thus, the snake is attracted very quickly and efficiently to the desired feature [11]. Elastic snake represented by a dataset of points such as n points $v_i$, where $i = 0,\ldots.n-1$, two energy terms $E_{internal}$ (internal elastic energy) and $E_{external}$ (external edge-based energy). Fitting of the contour on the image is controlled by $E_{internal}$ and $E_{internal}$ is used for maintaining the deformation. $E_{external}$ Consist $E_{image}$ and $E_{con}$ which represents image forces by image itself and constraint force that the user introduced respectively [12]. So, the Equation for the energy function is-

$$\begin{aligned} E^*_{snake} &= \int_0^1 E_{snake}(v(s))ds \\ &= \int_0^1 (E_{internal}(v(s)) + E_{image}(v(s)) \\ &\quad + E_{con}(v(s)))ds \quad (2) \end{aligned}$$

II. METHODOLOGY OF PROJECT

A. *Complexity Analysis of Trebeculae*

For the complexity quantifications of the trabeculae inside heart, we have used Fractal Dimension (FD) parameters that is described in introduction section. The block diagram for implementation of this section is shown in Fig. 3.

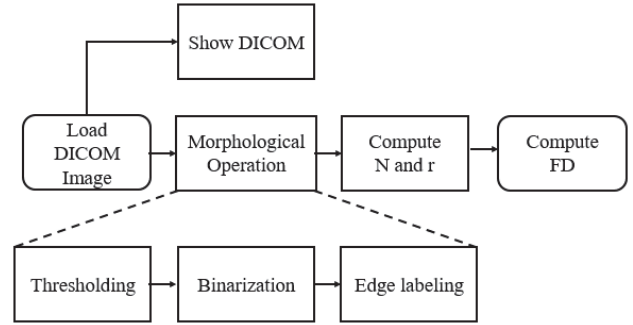

Fig. 3. Block presentation of complexity analysis

For the complexity analysis, we need some morphological operation on DICOM MRI to compute N is the number of boxes that cover the pattern, and r is the magnification. After computing N and r, the FD can be calculated by using Eq. 1. Layout of the Graphical User Interface (GUI) that is design for the complexity analysis of provided DICOM MRI of the heart is shown in Fig. 4. This layout in Fig. 4. Represents GUI for computing the FD. It's very simple and basic consisting DICOM image load and FD calculation button and after performing the operation it provide FD parameters.

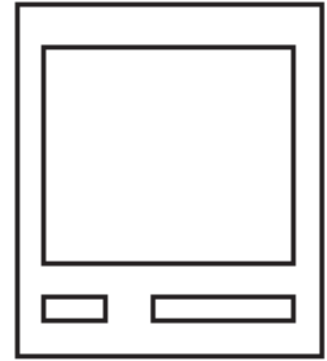

Fig. 4. Graphical User Interface layout for complexity analysis

## B. Manual Segmentation of Heart

For the complexity quantifications of the trabeculae inside heart, we have used Fractal Dimension (FD) parameters that is described in introduction section. The block diagram for implementation of this section is shown in Fig. 5.

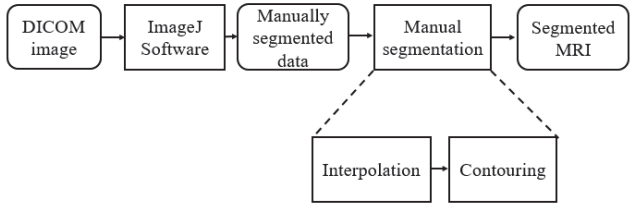

Fig. 5. Block presentation of manual segmentation

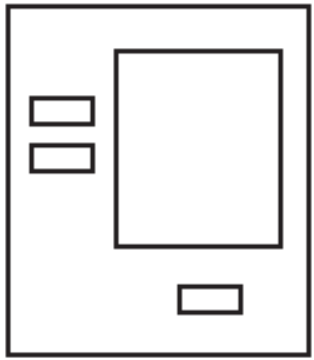

Fig. 6. Graphical User Interface layout for manual segmentation

From the block diagram in Fig. 5, it is required to manually select the contour and make a dataset using ImageJ software. After getting dataset, manual segmentation will be done which based on the interpolation for the contour from the given dataset. On the other hand, in Fig. 6, Layout of the implemented GUI for performing manual segmentation is shown that contains 2 input button for loading input manual segmented data and parameters and one execution button.

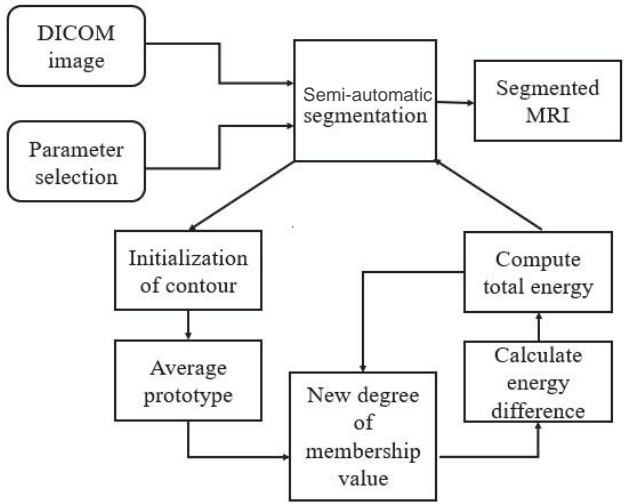

Fig. 7. Block presentation of semi-automatic segmenation

## C. Semi-automatic Segmentation of Heart

For the complexity quantifications of the trabeculae inside heart, we have used Fractal Dimension (FD) parameters that is described in introduction section. The block diagram for implementation of this section is shown in Fig. 7 which demonstrate the process of computing semi-automated segmentation. The process initiated by selecting DICOM image and parameter initialization. Afterward manual segmentation integrated into numbers of operation such as Initialization of contour, averaging prototypes, getting new membership values, calculating differences in energy computing total energy and using this new values as feedback.

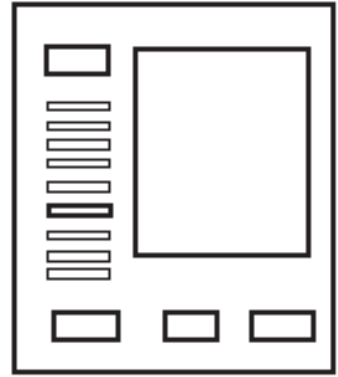

Fig. 8. Graphical User Interface layout for semi-automatic segmenation

Fig. 8 represents the layout of the implemented GUI for the finding the contour of MRI for the semi-automatic segmentation. This GUI has getter numbers of input and output blocks like, selecting image, defining iteration, selection and run buttons.

### III. RESULTS AND DISCUSSIONS

Overall project results can be divided into three separate section that are describe below-

### A. Results for Complexity Analysis of Trebeculae

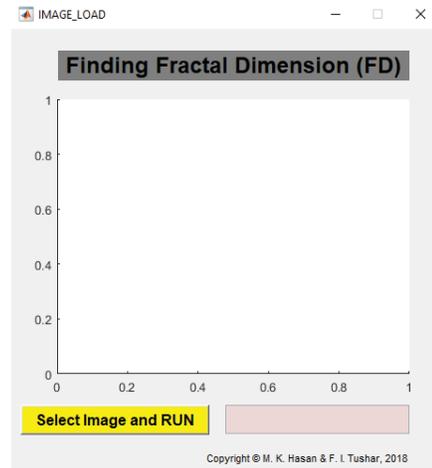

Fig. 9. Designed GUI for FD computation

Fig. 9 shows the implemented GUI for finding out the FD where by "**Select Image and Run**" button user selected the target DIOCOM image after that the process for calculating FD will be started. The FD indicates the amount of the complex geometric pattern of the inner part of the heart. The greater the value of the FD indicates the greater the complexity. For the given slice of MRI data, we see that the FD is 1.809 +/-0.16439 (Standard Deviation) as shown in Fig. 10. This result demonstrates the inner part of target MRI of the heart is quite complex.

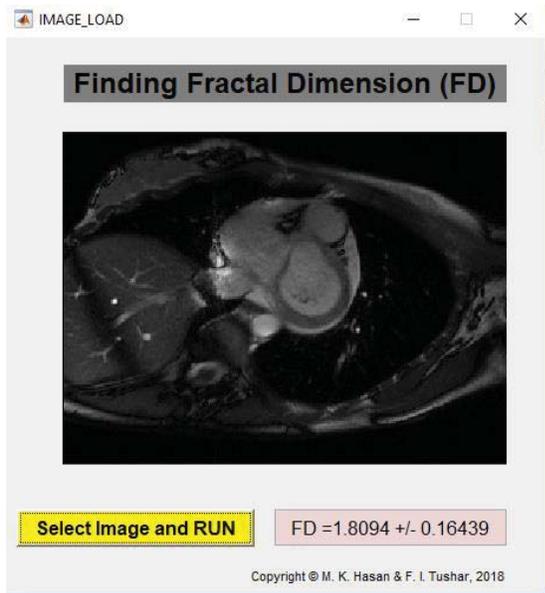

Fig. 10. Visualization of DICOM MRI and FD

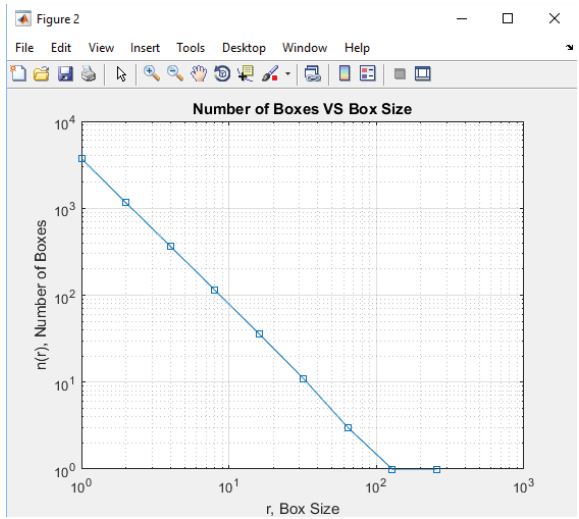

Fig. 11. Presentation of Number of boxes w.r.t box sizes

This results in Fig. 11 shows relation between the box scaling and the FD. If the scaling of the boxes increases then the boxes number containing information is decreased exponentially and FD is equivalent to this exponent. On the other hand, Fig. shows the natural logarithmic plot of the box sizes VS box number.

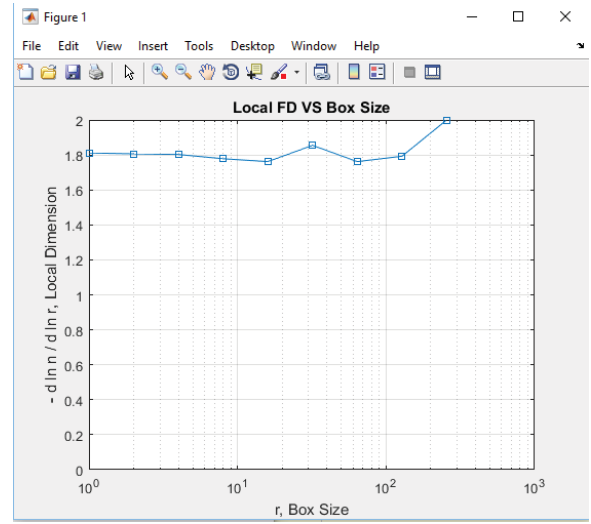

Fig. 12. Presentation of local dimension VS box sizes

*B. Result for Manual Segmentation of Heart*

The Graphical User Interface that is design for this sections is shown in Fig. 13. After loading DICOM MRI into the imageJ software, ROI is to be cropped for the manual segmentation then save as a .txt file. This processing repeated depending on the slice number.

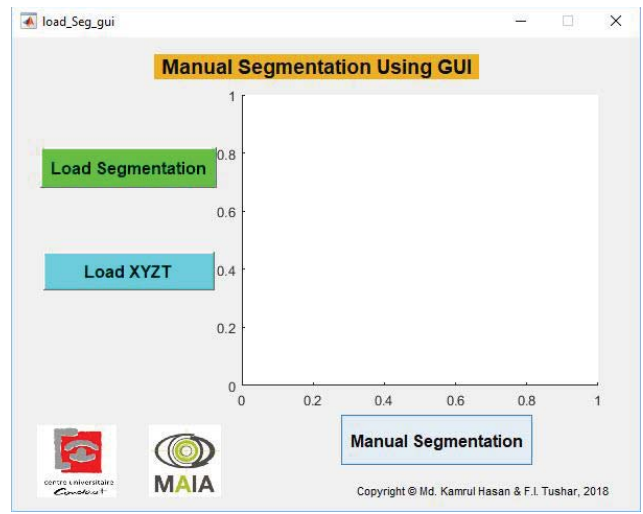

Fig. 13. Designed GUI for manual segmentation

After that, .txt data is been used in our implemented GUI to perform manual segmentation as shown in Fig. 14. This data is been in been interpolated and plotting the circle is been done. After all those processing segmented MRI is been achieved which shows the different cardiac border. The Green Circle shown the epicardium border of the heart, where red circle shown the endocardium border of the heart.

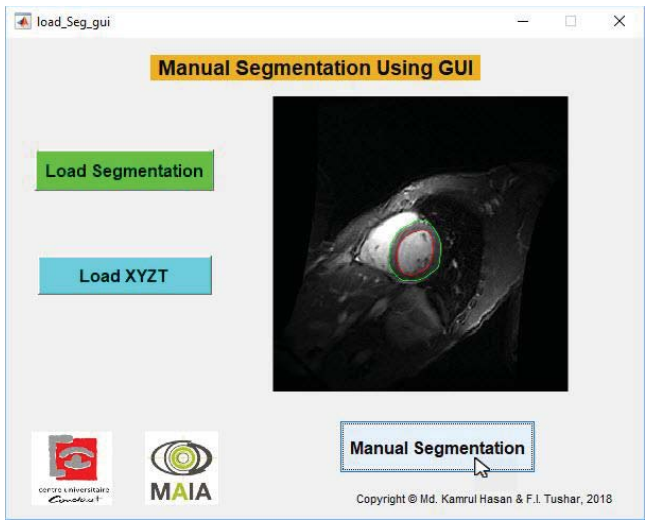

Fig. 14. Manually segmented epicardium and endocardium border

*C. Results for Semi-automatic Segmentation of Heart*

For the semi-automatic segmentation, the designed GUI is shown in Fig. 15. For the segmentation of MRI image using snake algorithm, we need to initialize the contour around the ROI then we need to iterate after defining some parameters required for the iteration. Initialization for the contour for the segmentation using snake algorithm is shown in Fig. 16. Then after selecting all the parameters that are mentioned in Fig. 16, we need to click run button. After some moments, it will provides a semi-automatic contour for the segmentation using snake algorithm as shown in Fig. 17. One of the vital parameters of this semi-automatic segmentation is time of running the program to make a contour. This executions time can be reduced by selecting the proper value of all the parameters that are mention in the GUI in Fig. 17.

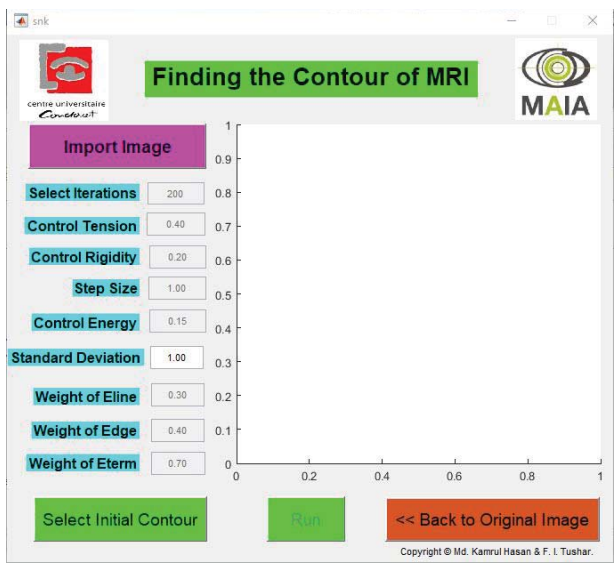

Fig. 15. Designed GUI for semi-automatic segmentation

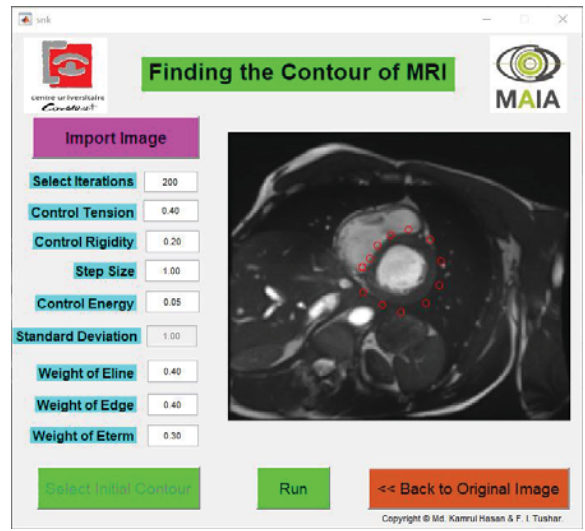

Fig. 16. Initialization of the contour for segmentation

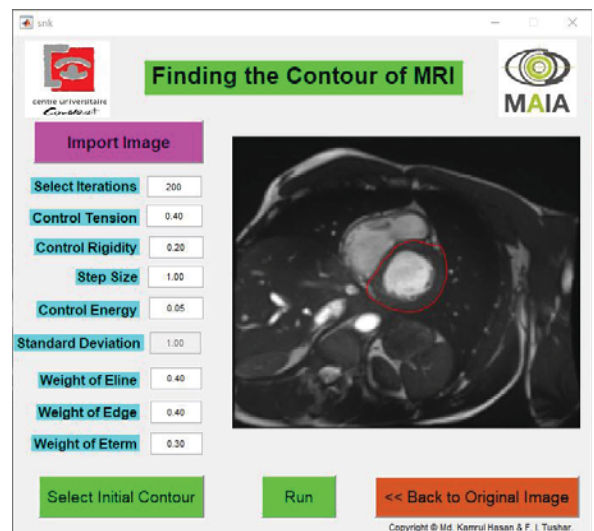

Fig. 17. Semi-automated segmentation

ACKNOWLEDGEMENT

We want to convey special gratitude to our project supervisor Dr. Alain LALANDE and all the course Teachers on Medical Sensor. Without their crucial guideline and teaching methodology, this project might be difficult to implement. We are really grateful to them.

REFERENCES

[1] P. A. Iaizzo, Handbook of Cardiac Anatomy, Physiology, and Devices, Spinger International Publishing Switzerland 2015.

[2] World Health Organization. Cardiovascular diseases (CVDs). [Accessed 9 Jan. 2018]. Available at:
http://www.who.int/mediacentre/factsheets/fs317/en/

[3] M. Niemann, S. Stork, and F. Weidemann, "Left Ventricular Noncompaction Cardiomyopathy: An Overdiagnosed Disease," Images in Cardiovascular Medicine.

[4] Fractals and Complexity. [Accessed 10 Jan. 2018]. Available at:

https://imagej.nih.gov/ij/plugins/fraclac/FLHelp/Fractals.htm

[5] G. Captur, V. Muthurangu, C. Cook, A. Flett, R. Wilson, A. Barison, D. Sado, S. Anderson, W. McKenna, T. Mohun, P. Elliott, and J. Moon, "Quantification of left ventricular trabeculae using fractal analysis," Journal of Cardiovascular Magnetic Resonance, vol. 15, no.36, 2013.

[6] S. Țălu, "Mathematical methods used in monofractal and multifractal analysis for the processing of biological and medical data and images," Animal Biology & Animal Husbandry, vol. 4, no. 1, pp.1-4, 2002.

[7] Fractal Foundation. [Accessed 10 Jan. 2018]. Available at: http://fractalfoundation.org/OFC/OFC-10-5.html

[8] C. Attikos and M. Doumpos, "Faster Estimation of the Correlation Fractal Dimension Using Box-counting," Fourth Balkan Conference in Informatics, Thessaloniki, pp. 93-95, 2009.

[9] X. Bresson, S. Esedoḡlu, P. Vandergheynst, J. Thiran, and S. Osher, "Fast Global Minimization of the Active Contour/Snake Model," Journal of Mathematical Imaging and Vision, vol. 28, no.2, pp.151-167, 2007.

[10] P. Makowski, T. Sorensen, S. Therkildsen, A. M. H. Stodkilde-Jorgensen, E. Pedersen, "Two-phase active contour method for semiautomatic segmentation of the heart and blood vessels from MRI images for 3-D visualization," Comput. Med. Imag. Graph., vol. 26, no. 1, pp. 9-17, 2002.

[11] A. Yezzi, S. Kichenassamy, A. Kumar, P. Olver and A. Tannenbaum, "A geometric snake model for segmentation of medical imagery," in IEEE Transactions on Medical Imaging, vol. 16, no. 2, pp. 199-209, April 1997.

[12] Dr. George Bebis, University of Nevada. [Accessed 11 Jan. 2018]. Available at: http://www.cse.unr.edu/~bebis/CS791E/Notes/DeformableContours.pdf